\documentclass[12pt]{elsarticle}
\biboptions{sort&compress}

\usepackage{amsmath,amsfonts,amssymb,amsthm, bm}
\usepackage{enumitem}
\usepackage{bm}
\usepackage{xspace}
\usepackage{collcell}
\usepackage{booktabs, siunitx}
\usepackage{color}
\usepackage{gensymb}
\usepackage{subfig}
\usepackage{placeins}
\usepackage[toc,page]{appendix}
\usepackage{multirow}

\def\Dz      {{\ensuremath{D^0}}\xspace}
\def\Dbar    {{\kern 0.2em\overline{\kern -0.2em D}{}}\xspace}
\def\Dzb     {{\ensuremath{\Dbar{}^0}}\xspace}
\def\KPIPIPI {\ensuremath{K^- \pi^+\pi^+\pi^-}}
\def\PIKPIPI {\ensuremath{K^+ \pi^-\pi^-\pi^+}}
\def\DCA     {\ensuremath{\Dz  \to \PIKPIPI}}
\def\CFA     {\ensuremath{\Dzb \to \PIKPIPI}}

\def\KSKPI   {\ensuremath{K^0_{\rm S} K^\pm\pi^\mp}}
\def\KS      {\ensuremath{K^0_{\rm S}}}
\def\xp      {\ensuremath{\mathbf{x}}}
\def\dx      {\mathrm{d}\mathbf{x}}
\def\CP      {\ensuremath{\mathcal{CP}\xspace}}
\def\deltaD  {\ensuremath{\delta_{K3\pi}}}
\def\deltaND {\ensuremath{\tilde{\delta}_{K3\pi}}}
\newcommand*{\mathhack}[1]{$#1$}
\newcommand{\mev}{\ensuremath{\mathrm{\,Me\kern -0.1em V}}\xspace}
\newcommand{\mevcc}{\ensuremath{{\mathrm{\,Me\kern -0.1em V\!/}c^2}}\xspace}

\journal{Physics Letters B}

\begin{document}

\begin{frontmatter}

\title{Improved sensitivity to the CKM phase $\gamma$ \\ through binning phase space in \\ $B^- \to DK^-$, $D \to K^+\pi^-\pi^-\pi^+$ decays}

\author[cern]{T.~Evans\corref{cor1}} \cortext[cor1]{Corresponding author}
\ead{timothy.david.evans@cern.ch}

\author[madras]{J.~Libby}
\author[oxford]{S.~Malde}
\author[oxford]{G.~Wilkinson}

\address[cern]{European Organisation for Nuclear Research (CERN), 1211 Geneva 23, Switzerland}
\address[madras]{Indian Institute of Technology Madras, Chennai 600036, India}
\address[oxford]{University of Oxford, Denys Wilkinson Building, Keble Road,  OX1 3RH, United Kingdom}

\begin{abstract}
A binning scheme is proposed for $D \to \PIKPIPI$ phase space that will improve the sensitivity of a $B^- \to DK^-$ analysis to the angle $\gamma$ of the Cabibbo-Kobayashi-Maskawa Unitarity Triangle.  The scheme makes use of amplitude models recently reported by the LHCb collaboration. Assuming that a four-bin scheme optimised on the models retains a similar sensitivity when applied in data, it is estimated that the statistical uncertainty on $\gamma$ from the $B$-meson sample so far collected by the LHCb experiment will be as low as 5 degrees. This will be one of the most precise results available for any single decay mode in a $B^- \to D K^-$ measurement. Quantum-correlated $D\bar{D}$ data accumulated by the CLEO-c experiment are analysed to provide first constraints on the coherence factors and average strong-phase differences in the four bins, which are necessary inputs for the measurement.  These constraints are compared with the predictions of the model, and consequences for the measurement of $\gamma$ are discussed.
\end{abstract}


\end{frontmatter}

\section{Introduction}
\label{sec:intro}
An important goal of flavour physics is to determine the angle $\gamma = {\rm arg}(-V_{ud}V^\ast_{ub}/V_{cd}V^\ast_{cb})$ of the Cabibbo-Maskawa-Kobayashi (CKM) Unitarity Triangle with the best possible precision.  Sensitivity to this weak phase (also denoted  $\phi_3$) can be obtained by measuring \CP-violating and associated observables in the decay $B^- \to DK^-$, where $D$ indicates a neutral charm meson reconstructed in a final state common to both $D^0$ and $\bar{D}^{0}$.  (The inclusion of charge-conjugate processes is implied throughout, unless otherwise stated.)  First measurements of $\gamma$ using this strategy were performed at the $B$-factory experiments~\cite{BABARGAMMA,BELLEGAMMA}, but the most precise ensemble of results now comes from  the LHCb collaboration, which has exploited a wide selection  of $D$-decay modes\footnote{Additional inputs to this result come from analogous strategies involving $B^0$ decays and from time-dependent $B^0_s$ and $B^0$ measurements.} to establish $\gamma = (74.0^{+5.0}_{-5.8})^\circ$~\cite{GAMMACONF}.
This value is consistent with, but significantly less precise than the prediction of $(65.6^{+1.0}_{-3.4})^\circ$, which comes from the knowledge of the sides and other angles of the Unitarity Triangle~\cite{CKMFITTER}, with a similar prediction in Ref.~\cite{UTFIT}.  Hence more data and new approaches are required to improve the precision of the direct measurement and to allow for more stringent tests of the CKM paradigm.

One input to the LHCb determination of $\gamma$ is the measurement of observables involving  $D \to \PIKPIPI$ decays, where this mode is reconstructed inclusively as part  of the $B^- \to DK^-$ decay chain.  Here there are contributions from both Cabibbo-favoured (CF) and doubly Cabibbo-suppressed (DCS) amplitudes in the charm-meson decay, as well as favoured $b \to c$ and suppressed $b \to u$ amplitudes in the $B$-meson decay,  the net effect of which is to introduce interference effects sensitive to $\gamma$~\cite{ADS}.   There are four possible decay configurations, with rates given by
\begin{eqnarray}
{\Gamma(B^{\mp}\to(K^{\pm}\pi^{\mp}\pi^{\mp}\pi^\pm)_D K^{\mp})} &  \propto &
 (r_{B})^{2} + (r_{D}^{K3\pi})^{2}  +  \nonumber \\
& & 2r_{B}r_{D}^{K3\pi}R_{K3\pi}\cos{(\delta_{B}+\deltaD \mp \gamma)}\,  \label{eq:adssuppressed} \\
{\Gamma(B^{\mp}\to(K^{\mp}\pi^{\pm}\pi^{\pm}\pi^\mp)_D K^{\mp})} &  \propto &
 1 + (r_B r_{D}^{K3\pi})^{2}  +  \nonumber \\
& & 2r_{B}r_{D}^{K3\pi}R_{K3\pi}\cos{(\delta_{B}-\deltaD\mp \gamma)}\,  \label{eq:adsfavoured},
\end{eqnarray}
where $r_B \sim 0.1$ is the absolute ratio of $B^- \to {\bar D}^0 K^-$  to $B^- \to D^0 K^-$ amplitudes.
The phase difference between these two amplitudes is $(\delta_B - \gamma)$, where $\delta_B$ is a \CP-conserving strong phase.   (These expressions neglect small corrections from $D^0\bar{D}^0$ mixing, which can be included in a straightforward manner~\cite{RAMA}.)

The other parameters in Eqs.~\ref{eq:adssuppressed} and~\ref{eq:adsfavoured}  relate to  the phase-space averaged contribution of intermediate hadronic resonances in the charm decay, which are in general different for the CF and DCS amplitudes. In particular the coherence factor  $R_{K3\pi}$, which takes a value between 0 and 1, and average strong-phase difference $\deltaD$~\cite{ATWOODSONI} are defined by
\begin{equation}
 R_{K3\pi}e^{i\deltaD}  = 
  \frac{\int\mathcal{A}_{\CFA}(\xp)\mathcal{A}^\ast_{\DCA}(\xp)\dx }{A_{\CFA} A_{\DCA} } ,
\label{eq:coherence}
\end{equation}
where $\mathcal{A}_{\Dz(\Dzb)\to\PIKPIPI}(\xp)$ is the decay amplitude of $\Dz(\Dzb)\to\PIKPIPI$ at a point in multi-body phase space described by parameters $\xp$, and 
\begin{equation}
  A_{\Dz(\Dzb)\to\PIKPIPI}^{2}=\int |\mathcal{A}_{\Dz(\Dzb)\to\PIKPIPI}(\xp)|^{2}\dx.  
\label{eq:amplitude}
\end{equation}
Therefore $A_{\CFA}$ is the CF amplitude, averaged over phase space, and  $A_{\DCA}$ is the corresponding DCS quantity.  The mean ratio of suppressed-to-favoured decay amplitudes is
\begin{equation}
  r_{D}^{K3\pi} =  \frac{A_{\DCA}}{A_{\CFA}}.
\label{eq:rd}
\end{equation}
Throughout the discussion the approximation is made that \CP-violation can be neglected in the charm system~\cite{HFLAV}, and expressions are given in the convention $\CP|D^0\rangle = |\bar{D}^0\rangle$. 

The phase-space averaged hadronic parameters defined in Eqs.~\ref{eq:coherence} and~\ref{eq:rd} may be accessed both from quantum-correlated $D\bar{D}$ decays at the $\psi(3770)$, such as in the data set of the CLEO-c experiment and through studies of $D^0\bar{D}^{0}$ mixing~\cite{EVANS, SAM,LHCBMIX}.  A combination of measurements made from both sources yields $R_{K3\pi} = 0.43^{+0.17}_{-0.13}$, $\deltaD = (128^{+28}_{-17})^\circ$ and $r_D^{K3\pi} = (5.49 \pm 0.06) \times 10^{-2}$~\cite{EVANS}.  Noting these values and inspecting Eq.~\ref{eq:adssuppressed}, it can be seen that the \CP-violating effects in $B^\mp \to D K^\mp$, $D \to K^{\pm}\pi^{\mp}\pi^\mp\pi^\pm$ decays are expected to enter at leading order, whereas Eq.~\ref{eq:adsfavoured} indicates that there will be negligible \CP\ violation in $B^\mp \to D K^\mp$, $D \to K^{\mp}\pi^{\pm}\pi^\pm\pi^\mp$ decays.  These predictions have been confirmed by LHCb; in particular a \CP\ asymmetry of $-0.31 \pm 0.11$ is measured for the former pair of modes~\cite{LHCBGAMK3PI}.

Although $D \to \PIKPIPI$ has a significant weight in the global LHCb determination of $\gamma$ with $B^- \to DK^-$ decays, the inclusive nature of the analysis brings limitations that prevent the full power of this mode from being harnessed.  The integration over the phase space of the $D$ decay dilutes the quantum interference, which is manifested in the fact that the coherence factor is much smaller than unity.   A more attractive approach is to perform the analysis in disjoint bins of phase space.  In this case the parameters of Eqs.~\ref{eq:coherence},~\ref{eq:amplitude} and~\ref{eq:rd} are re-defined within each bin.  A given bin, if suitably chosen, can possess greater coherence and therefore exhibit enhanced interference effects than the integral over all phase space. 
In addition to improving the intrinsic sensitivity, this strategy also has the benefit of breaking degeneracies that exist in the inclusive analysis, thus enabling a single solution to be obtained from the data  in the physical region of the Unitarity Triangle plane.

The choice of a performant binning scheme requires good knowledge of the variation of CF and DCS amplitudes across the phase space.  Although amplitude models had been constructed for the decay $D^0 \to \KPIPIPI$~\cite{MARKIIIK3PI,BESIIIK3PI}, until recently none existed for the suppressed $D^0 \to \PIKPIPI$ mode. Earlier attempts~\cite{SAM2} to define a binning scheme have been hindered by this lack of knowledge. However, the large charm data set collected by LHCb has allowed this omission to be rectified. A recent publication~\cite{LHCBK3PI} reported the world's first amplitude model of  $D^0 \to \PIKPIPI$ and a new model of the favoured $D^0 \to \KPIPIPI$ mode that benefits from a much larger sample  than available to any previous study.   In principle these models can be used directly in an unbinned $B^- \to DK^-$ measurement, thereby maximising the statistical precision of the analysis.  However, this strategy has the drawback that any imperfections in the models have the potential to bias the $\gamma$ determination in a manner that is difficult to assess.   Instead, it is preferable to use the models to define a set of bins with a good statistical sensitivity to $\gamma$. The hadronic parameters of the $D$ decay can then be measured directly within each bin, either from threshold or $D^0\bar{D}{}^0$-mixing data, or a combination of both. The impact of any shortcomings of the models will be merely to reduce the statistical sensitivity of the analysis with respect to expectation, leaving the measured value of $\gamma$ unbiased and model independent.   A similar philosophy has been advocated and followed for self-conjugate decays such as $D \to K^0_{\rm S}\pi^+\pi^-$, $D\to K^0_{\rm S}K^+K^-$ \cite{GGSZ, BONDAR, CLEOKSPIPI, LHCBKSHH, LHCBKSHH2, BELLEKSHH, BELLEKSHH2} and $D \to K^0_{\rm S}\pi^+\pi^-\pi^0$ \cite{KSPIPIPI0_CLEO, KSPIPIPI0_BELLE}.

The remainder of this Letter is organised as follows.  In Sec.~\ref{sec:binning} the LHCb amplitude models of $D^0 \to K^\pm \pi^\mp \pi^\mp\pi^\pm$ decays are introduced, and these models are then employed to determine a binning scheme that is predicted to have greatly improved sensitivity to $\gamma$ in a $B^- \to DK^-$ analysis.  In Sec.~\ref{sec:cleoc} measurements of the hadronic parameters of the $D$ decay are presented for the binning scheme, obtained from data collected by the CLEO-c experiment.
Conclusions are drawn in Sec.~\ref{sec:conclusions}.

\section{Binning scheme}
\label{sec:binning}
In this section, the LHCb models are reviewed, and then a binning scheme proposed to maximise sensitivity to $\gamma$.  The expected sensitivity of this scheme is then presented.

\subsection{The LHCb amplitude models}

The decay modes  $D^{0} \to K^{\pm} \pi^{\mp}\pi^{\mp} \pi^{\pm}$ have been studied by the LHCb collaboration and discussed in Ref.~\cite{LHCBK3PI}, which reports amplitude models for both final states.
The largest contributions to each decay mode are found to be via external $W$-emissions, 
manifesting themselves as the axial meson $a_1(1260)^{-}$ in the favoured mode and the axial kaons $K_1(1270)^{+}$ and $K_1(1400)^{+}$ in the suppressed mode. 
Large contributions are also found in both decay modes from vector-vector, {\it e.g.} $K^*(892)^{0}\rho(770)^0$,  and scalar-scalar amplitudes.
The values of the average strong-phase differences and coherence factors in local regions of phase space, as predicted by the model, are found to have small variations with respect to the choice of model components and parameters. Depending on the region of phase space considered, the predicted strong-phase differences and coherence factors are found to vary within a range of $\left[1^\circ, 7^\circ\right]$ and $\left[0.005, 0.030\right]$, respectively. This stability gives confidence when using these models to determine model-independent binning schemes.

\subsection{Bin definitions}

The phase space of the $D$-meson decay is divided into disjoint regions using the model predictions for the strong-phase difference between the favoured and suppressed amplitudes.  The models are only sensitive to the variation in the strong-phase difference across phase space.  Hence it is necessary to define a {\it normalised phase difference}, $\deltaND$, which at a position in phase space $\textbf{x}$ is given by
\begin{equation}
  \begin{split}
    \deltaND(\xp) &= \mathrm{arg}\left( \mathcal{A}_{\CFA}(\xp) \mathcal{A}^{\ast}_{\DCA}(\xp) \right) \\ 
                                &- \mathrm{arg}\left( \int \mathcal{A}_{\CFA}(\xp^\prime) \mathcal{A}^{\ast}_{\DCA}(\xp^\prime) \dx^\prime \right).
  \end{split}
\end{equation}
Hence the normalised phase difference averaged over all phase space is equal to zero by construction. 
In order to obtain the model prediction of the absolute phase difference in each bin, it is necessary to add to $\deltaND$ the measured value of the average phase difference $\deltaD$, as defined in Eq.~\ref{eq:coherence}.

\begin{figure}[t!]
  \centering
  \subfloat[S][]{\label{fig1a}\includegraphics[width=0.49\textwidth]{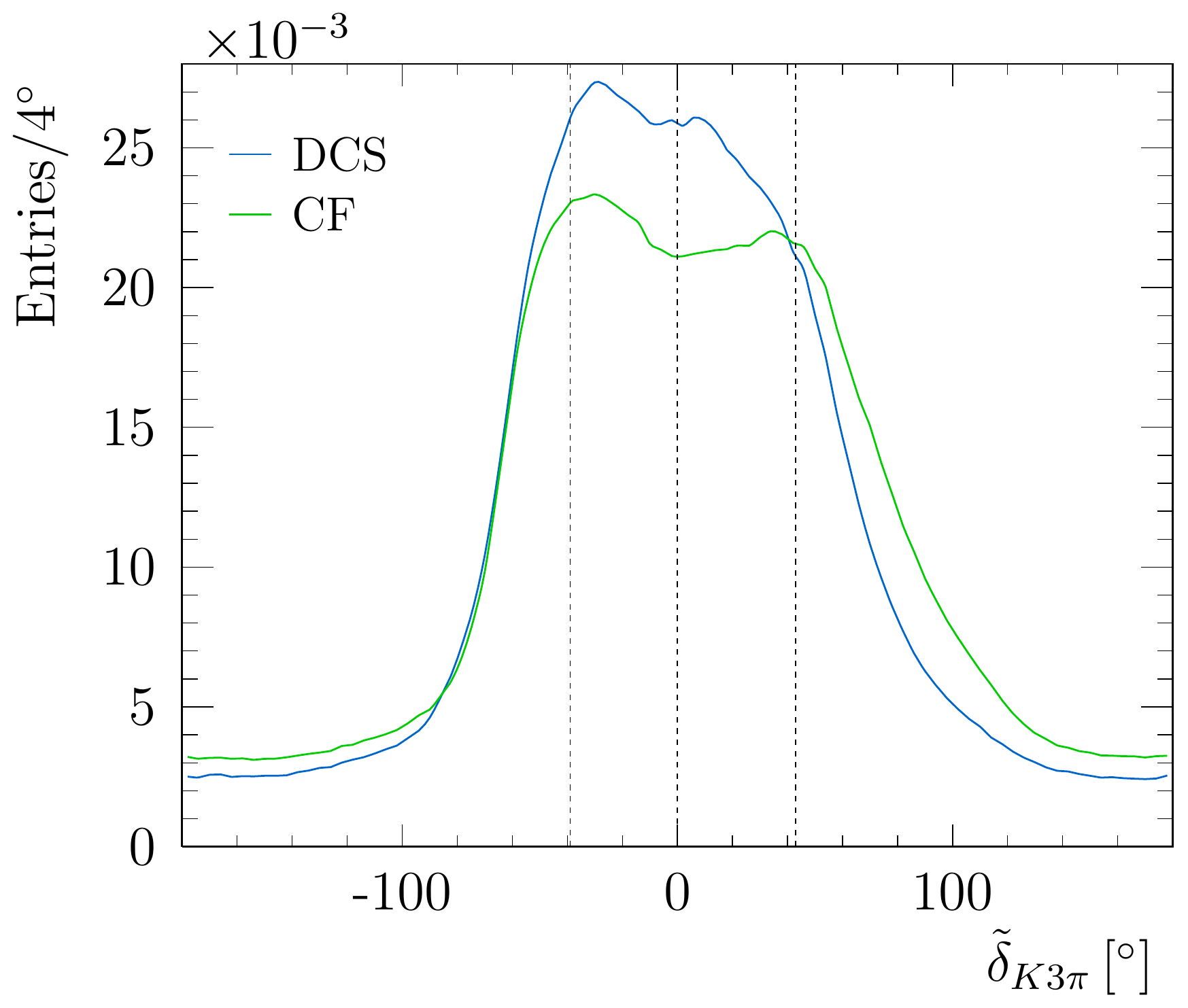}}
  \hspace{1.5mm}
  \subfloat[S][]{\label{fig1b}\includegraphics[width=0.49\textwidth]{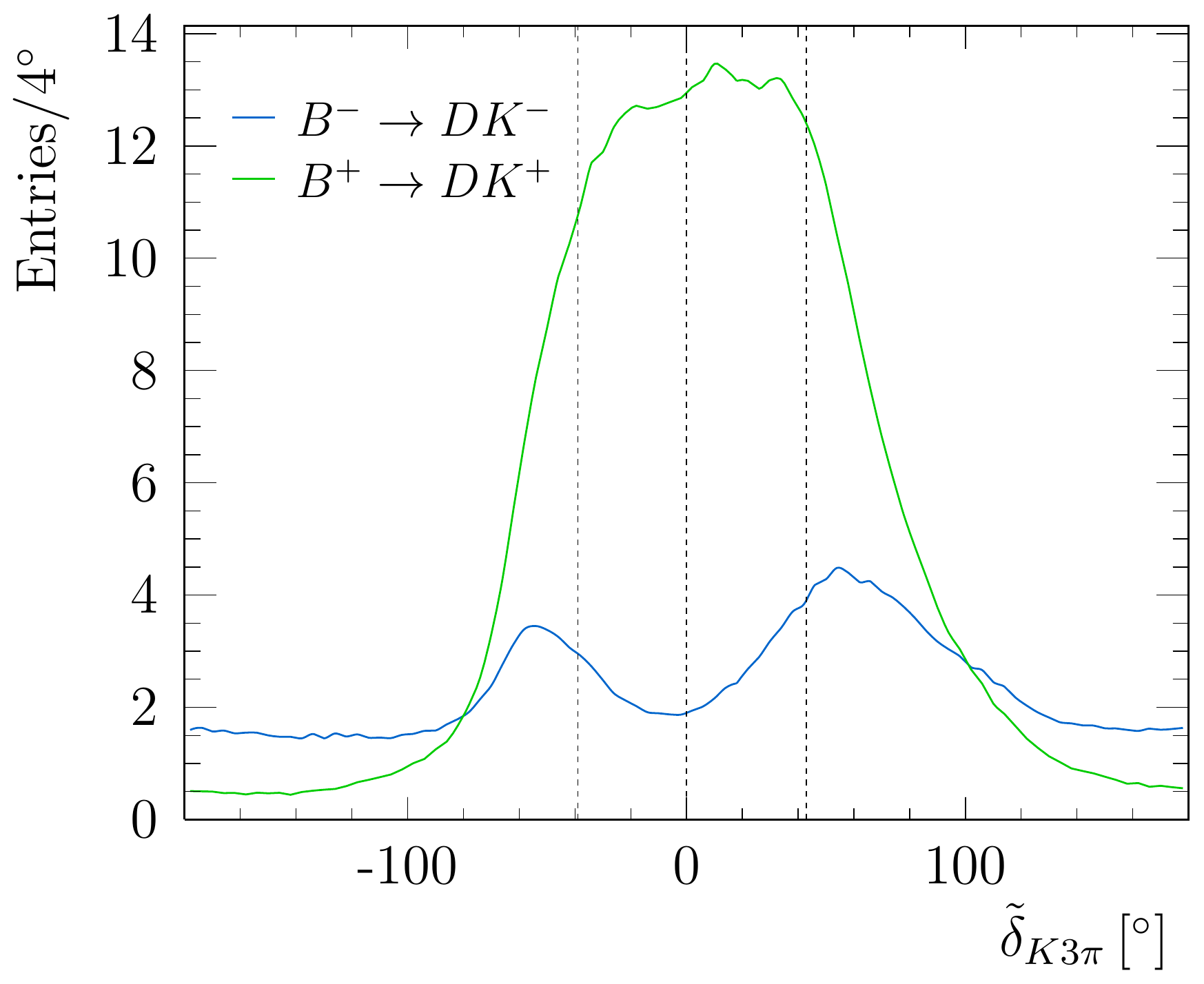}}
  \caption{\label{fig:phaseDifference}Distribution of the normalised phase difference of simulated $D \to K^+ \pi^- \pi^- \pi^+$ decays. 
         Shown in (a) are the expected distributions for the pure DCS and CF amplitudes, 
         where each of the histograms has been normalised to unit area.
         Shown in (b) is the expected distributions for $B^\mp \to D\left[K^\pm\pi^\mp\pi^\mp\pi^\mp\right] K^\mp$, 
         normalised to 600 $B$-meson decays.  
         Dashed vertical lines indicate the boundaries of the four bins discussed in the remainder of this paper.}
\end{figure}

Two simulations of very large samples are performed in order to visualise how the coordinate $\deltaND$ is distributed over the four-body phase space: one assuming an amplitude that follows the CF model, and one that follows the DCS model of LHCb.
Histograms of the normalised phase difference of each simulated decay are plotted for each scenario in Fig.~\ref{fig1a}. 
Shown in Fig.~\ref{fig1b} are the expected distributions of normalised phase difference for $B^\pm$ decays, assuming the values of $\gamma$ and the hadronic parameters from Ref.~\cite{GAMMACONF}. 
The histograms in the latter plot are normalised such that the yield summed over the $B$-meson charges corresponds to about 600 candidates, which is approximately the size of the signal sample that can be reconstructed in the full LHCb Run~1 and Run~2 sample of $9\,{\rm fb^{-1}}$, estimated by scaling the yields reported in the $3 \,{\rm fb^{-1}}$ Run 1 analysis~\cite{LHCBGAMK3PI} and accounting for the change in $b$-production cross-section between Runs~1~and~2. 

In order to arrive at a sensitive binning scheme, the bins should be well separated in average strong-phase difference, which is achieved by dividing the phase space with a one-dimensional binning in the normalised strong-phase difference.
This choice also results in bins with higher coherence than integrating over the phase space, which further improves the sensitivity.
Similar considerations apply when choosing a binning scheme for the self conjugate decays $D^0 \to K^0_{\rm S, L} \pi^+\pi^-$ and $D^0 \to K^0_{\rm S, L}K^+K^-$ \cite{BONDAR, CLEOKSPIPI}.
In order for each bin to have a similar weight in the analysis, and therefore benefit the overall precision, it is also necessary to distribute approximately equally the expected number of $\mathcal{CP}$-averaged decays in each bin.   Therefore, it is required that the different bins have an equal product of the CF and DCS phase-space averaged amplitude integrals:
\begin{equation}
  \mathcal{I}_i^2 = A^i_{\CFA} A^i_{\DCA}.
  \label{eq:Ii}
\end{equation}
This choice results in an approximate equipartition of a $B$ dataset between the different bins, which can be seen in Fig.~\ref{fig1b}. 
A more complex phase-space binning scheme, based on minimising the expected uncertainty on $\gamma$, results in a negligible improvement in sensitivity, when accounting for the uncertainties on the input parameters of the procedure. Hence the simpler division based upon Eq.~\ref{eq:Ii} is used.

The procedure to calculate the normalised phase difference as a function of position in the phase space is provided by Ref.~\cite{K3PICODE}, in addition to the definitions of the different bins, and examples of how to apply the binning scheme.

\subsection{Expected sensitivity}
\label{subsec:toys}

\begin{figure}[t!]
  \centering
  \subfloat[S][]{\label{fig2a}\includegraphics[width=0.49\textwidth]{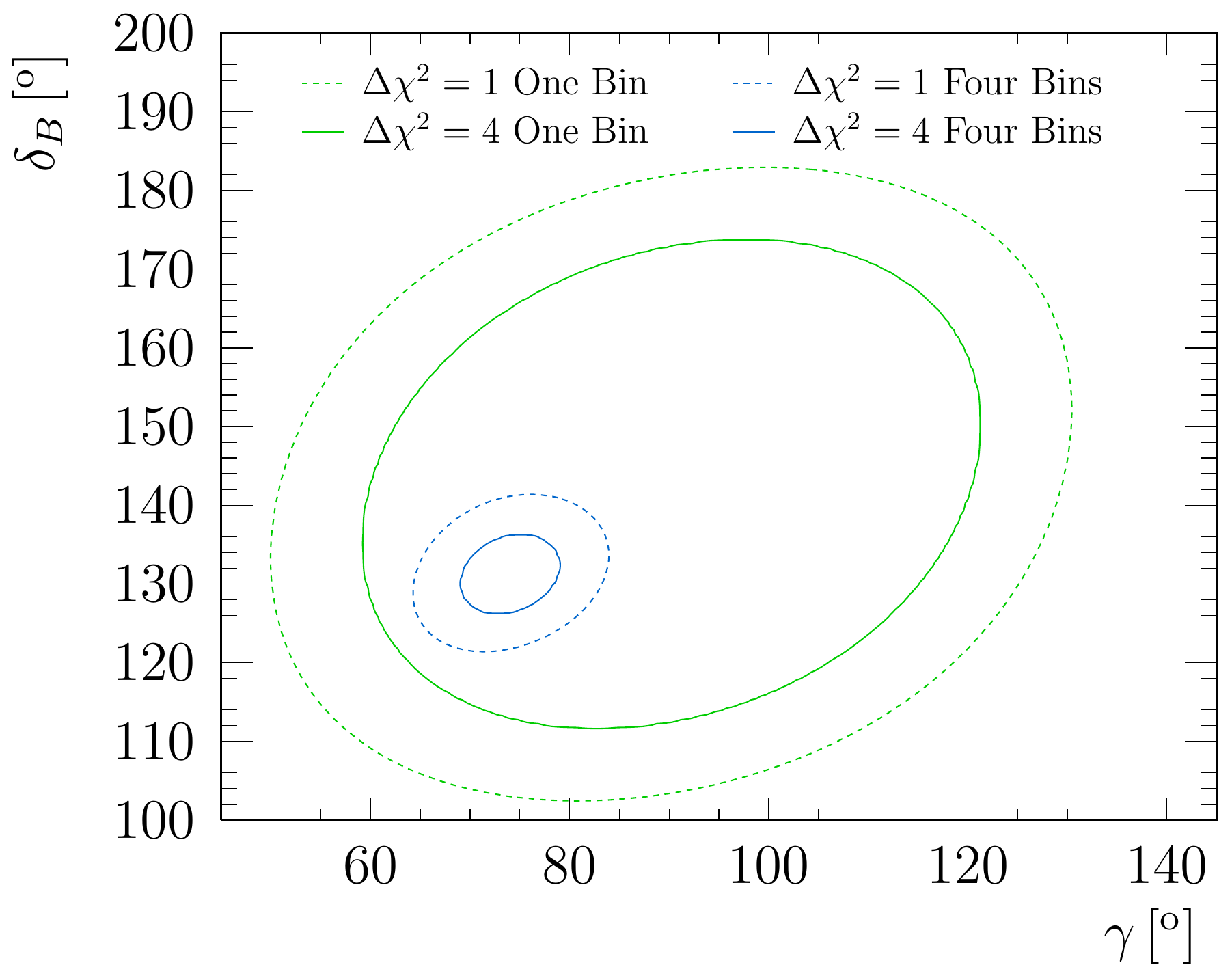}}
  \hspace{1.5mm}
  \subfloat[S][]{\label{fig2b}\includegraphics[width=0.49\textwidth]{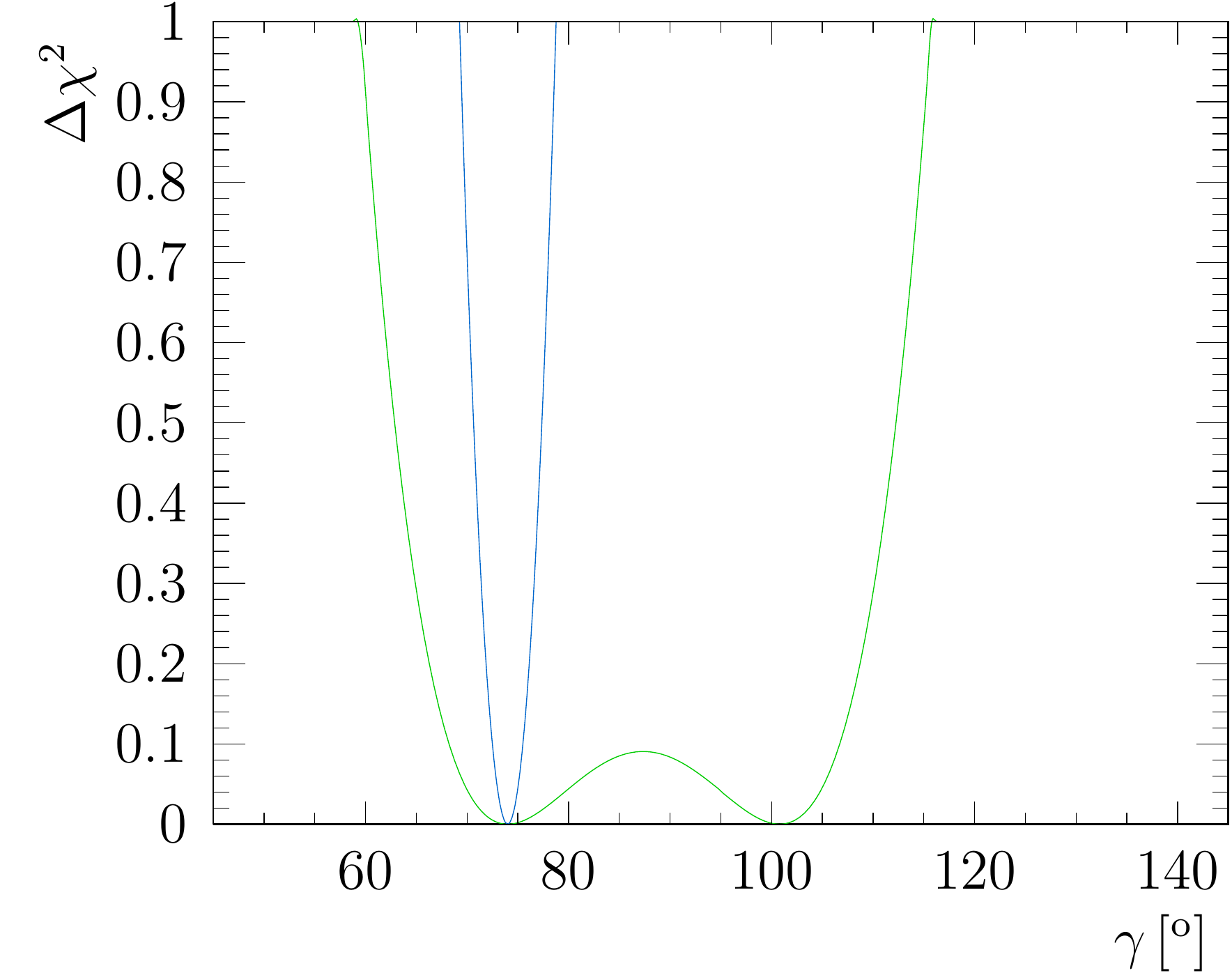}}
  
  \subfloat[S][]{\label{fig2c}\includegraphics[width=0.49\textwidth]{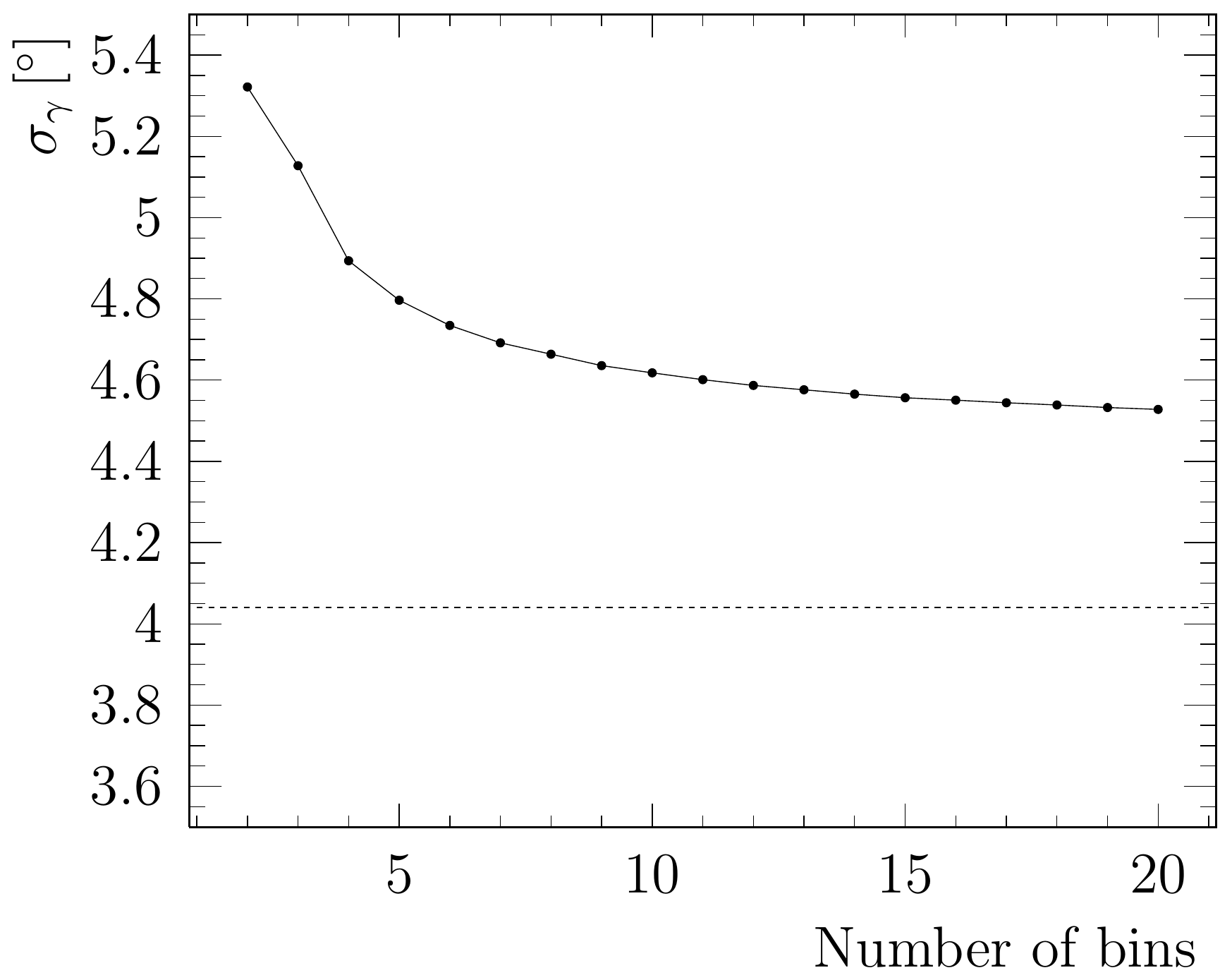}}
  \caption{\label{fig2} (a): Expected uncertainty contours in the $\gamma$ vs. $\delta_B$ plane for one and four bins from around 600 $B$ decays. (b): One-dimensional slice of the $\Delta \chi^2$ in $\gamma$ at $\delta_B = 131^\circ$, showing the very weakly separated double minima for the one-bin case. 
 (c): The expected sensitivity to $\gamma$ vs the number of bins in the phase space. The dashed line indicates the expected sensitivity from a unbinned model-dependent analysis. 
  }
\end{figure}

Simulated pseudoexperiments are performed in which  ${B^-  \to D K^-}$, $D\to \PIKPIPI$ data sets are generated and the $D$ mesons decayed according to the LHCb models with a uniform acceptance. Each pseudoexperiment comprises around $600$ decays. The LHCb Run~1 analysis achieved high purity and so no background contribution is included. 
Each simulated data set is fitted to determine $\gamma$, as well as the auxiliary parameters $r_B$ and $\delta_B$.  The fits are made in the context of a binned analysis, for differing numbers of bins, and also an unbinned analysis using the amplitude models.  The input values for $\gamma$ and the auxiliary parameters are taken from Ref.~\cite{GAMMACONF}.  In the fits, perfect knowledge is assumed of the properties of the $D$-meson decays, which in the case of the binned analysis are the local hadronic parameters, and  for the unbinned case are the amplitude models.

The expected contours in the $\gamma$ vs. $\delta_B$ plane are shown in Fig.~\ref{fig2a}, comparing four bins in phase space against the expected contours without dividing the phase space, {\it i.e.} a single bin. 
The contours are much wider in the single bin case, which is due both to a significantly worse intrinsic sensitivity around the true solution, and also the presence of an overlapping second minimum in $\gamma$. 
These features can be seen in Fig.~\ref{fig2b}, which shows a one-dimensional slice of the $\chi^2$ in $\gamma$ at $\delta_B$ at the expected value of $131^\circ$. 
The four-bin case does not suffer from the same degeneracy and thus has a single, narrow minimum. 
In fact, even a partitioning of the phase space into only two bins is sufficient to break the aforementioned degeneracy, thus providing a significantly improved precision compared to the single-bin case. 

The uncertainty as a function of the number of bins is shown in Fig.~\ref{fig2c} for two bins and above. 
As expected, the precision of the binned analysis gradually improves as the number of bins increases, and (within the range explored) saturates at about twelve bins, with around  a 10\% degradation  compared to the unbinned model-dependent method.
The size of the currently available $\psi(3770)$ data sets restricts the number of bins for which hadronic parameters can be determined.  The CLEO-c data set, for which results are presented in Sec.~\ref{sec:cleoc}, does not allow a stable analysis for more than three or four bins. The choice of four bins is made in light of the anticipated analysis of the larger sample that has been collected by BESIII. With four bins the statistical uncertainty on $\gamma$ from the $B$-meson data alone is expected to be $4.9^\circ$.
 
\section{Measurements of the hadronic parameters with CLEO-c data}
\label{sec:cleoc}
\subsection{Analysis of the $\psi(3770)$ data set}

A data set of $e^{+}e^{-}$ collisions produced by Cornell Electron Storage Ring  at $\sqrt{s} = 3.77\,\mathrm{GeV}$, corresponding to   $818\,\mathrm{pb}^{-1}$ of integrated luminosity and collected with the CLEO-c detector, is analysed to obtain first constraints on the hadronic parameters of $D \to \PIKPIPI$ decays in bins of phase space.  The analysis relies on the quantum-correlated nature of the $D\bar{D}$ pairs produced in the $\psi(3770)$ decay, and proceeds through counting the yields of various double-tagged events, where one $D$ meson is reconstructed in the signal mode $D \to \PIKPIPI$ and the other in a tag mode, for example a $\CP$ eigenstate ({\it e.g.} $D \to K^+K^-$, $D \to K^0_{\rm S} \pi^0$ {\it etc.}).  Full details on how the rates of each class of double-tagged event may be related to the hadronic parameters can be found in Ref.~\cite{EVANS}, and references therein.

The same set of tags are employed as in the phase-space integrated analysis~\cite{EVANS}. The selection criteria and procedures used to determine the background contributions are in common with the earlier study. In addition, the background contribution from $D \to \KSKPI $ decays  is explicitly removed by vetoing events containing signal candidates with a $\pi^+\pi^-$ invariant mass lying within $\pm10\mev$ of the nominal $\KS$ mass.

The division of candidates between multiple phase-space bins results in many of the individual yields being very low. For this reason the raw yields themselves are used as observables to determine the hadronic parameters in a log-likelihood fit.  Poisson terms are included of the form
\begin{equation}
  \log\left( \frac{\mu^{n} e^{-\mu}}{n!} \right) = n \log\mu - \mu - \log{n!},
\end{equation}
where $n$ is the observed number of candidates in a given bin of a given tag, and $\mu$ is the predicted number of candidates.
The coherence factor in each bin is required to be within the physical region, that is $0 \leq R_{K3\pi} \leq 1$, such that the expected signal yield is always positive definite and hence the likelihood well-defined. 
The predicted number of background candidates, as well as normalisation factors and efficiency corrections, are therefore accounted for with their relevant uncertainties in the expected number of candidates. 

The different categories of double tags, and the number of observables they bring to the analysis for $N$ bins of phase space, are as follows: 
\begin{description}
  \item[$\CP$ tags] Eleven $\CP$-eigenstate tags contribute $11 \times N$ observables;
  \item[Like-signed-kaon tags] The two tags, $K^+ \pi^-$  and $K^+ \pi^- \pi^0$, each contribute $N$ observables for those double tags where the charge of the kaon is the same in the signal and tag mode. (Those double tags where the two kaons are of opposite charge carry negligible sensitivity to the hadronic parameters and are used for normalisation purposes.); 
  \item[Self tags] The self tags, where both signal and tag are $\PIKPIPI$ with identical kaon charge (again, opposite charge double tags are used for normalisation), contribute $N^2$ yields as both sides of the decay are associated with a bin of phase-space.
    The number of distinct observables is however $N(N+1)/2$, as the signal side of the decay and tag side of the event can be freely exchanged;
  \item[Self-conjugate tags] The self-conjugate final state $\KS\pi^+\pi^-$ is divided into the 16 bins according to the `equal $\Delta \delta_D$' scheme described in Ref.~\cite{CLEOKSPIPI}. Hence for $N$ bins on the signal side, there are $16\times N$ observables.  
\end{description}
The total number of yields included in the fit is therefore $29N + \frac{N}{2}(N+1)$, which for the case of four bins is $126$.

Terms involving an equivalent set of double tags involving unbinned $D \to K^+\pi^-\pi^0$ decays as signal, with the yields  reported in Ref.~\cite{EVANS}, are included in the log-likelihood function to constrain the hadronic parameters of that system. Finally, terms involving $D \to K^+\pi^-$  decays tagged with $\CP$ eigenstates are also added to provide normalisation, again following the same procedure as in the earlier analysis. 
The branching fractions and other parameters used in Ref.~\cite{EVANS} are included in the likelihood as external constraints, including the phase-space averaged ratio of branching fractions and
global mixing results from Ref.~\cite{SAM}. 
Systematic uncertainties are assigned using the procedure followed in the earlier analysis~\cite{EVANS}.

Information on the binned double-tag results expressed in terms of the $\Delta$ and $\rho$ observables defined in Ref.~\cite{EVANS} and the yields of the self-conjugate tags can be found in appendix.

\begin{figure}
  \centering 
  \includegraphics[width=0.495\textwidth]{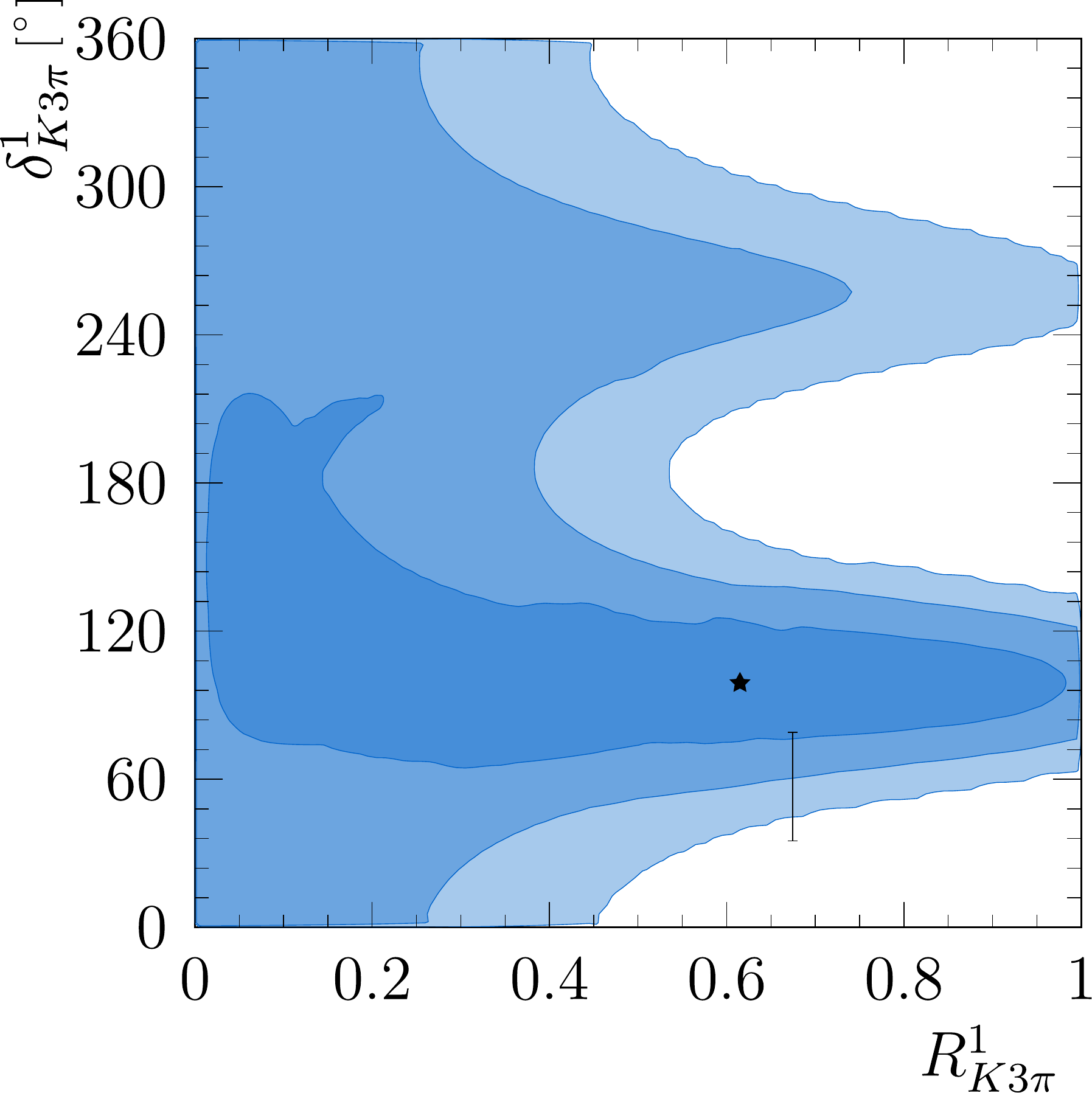} \includegraphics[width=0.495\textwidth]{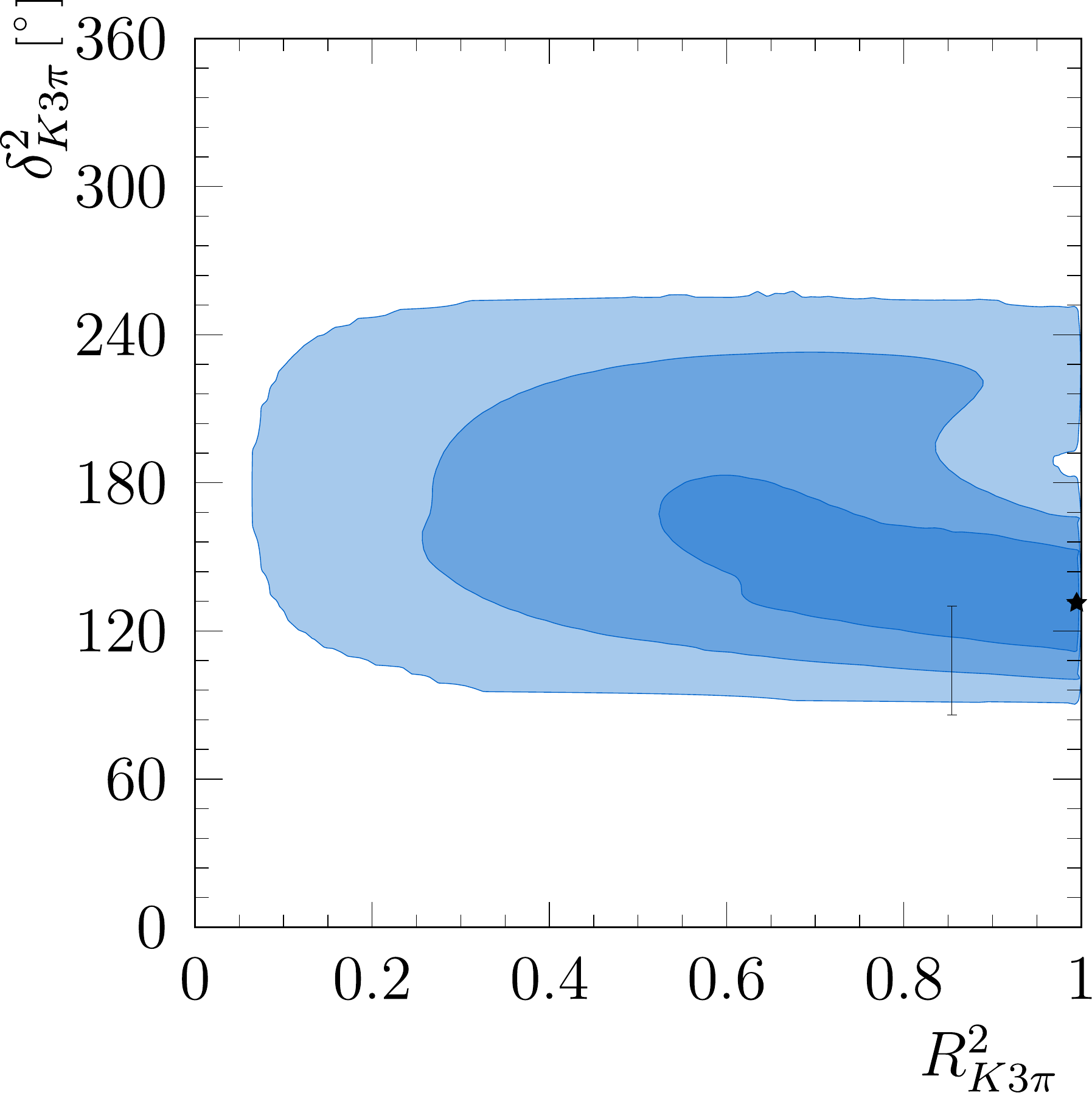}

  \includegraphics[width=0.495\textwidth]{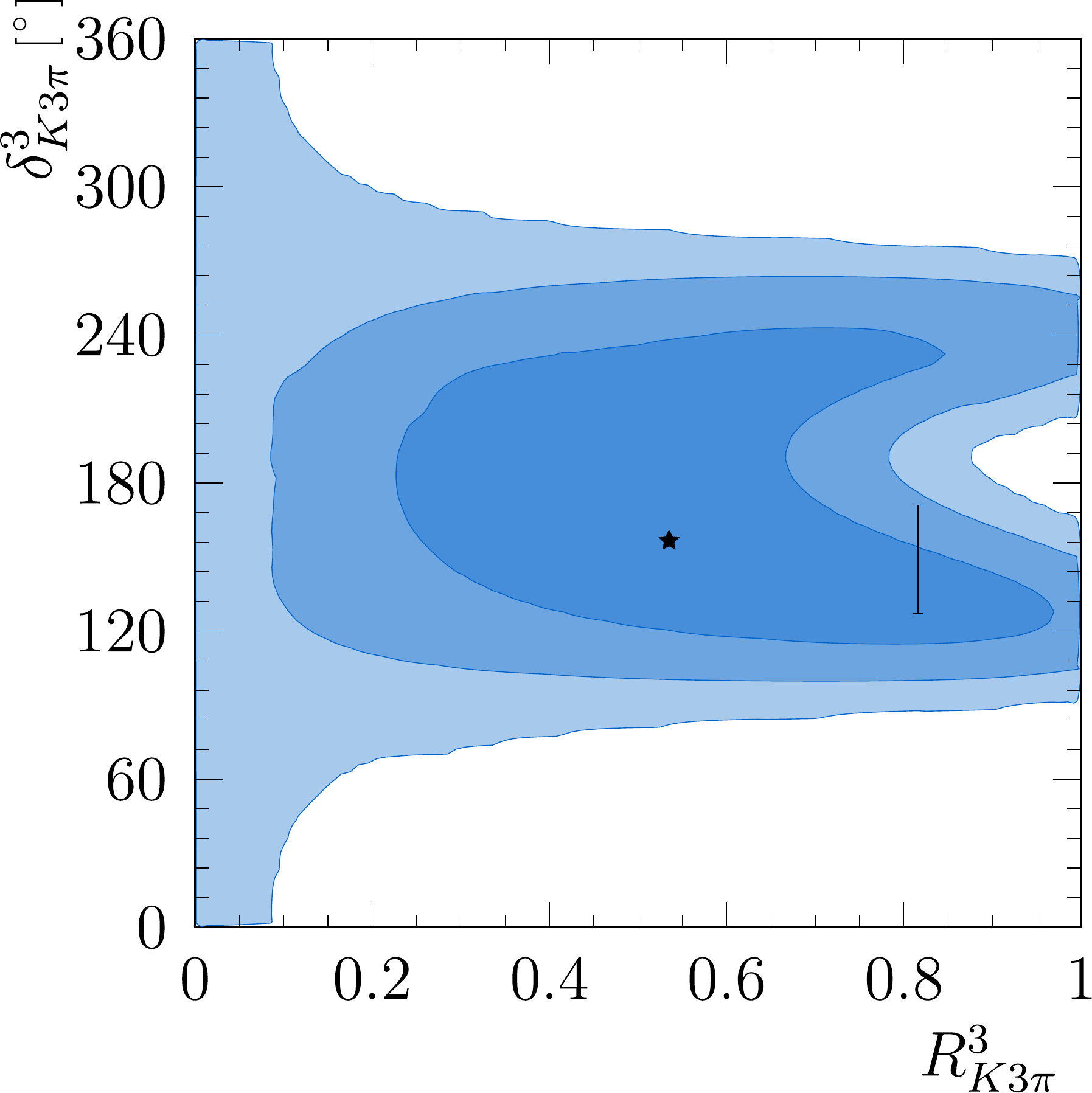} \includegraphics[width=0.495\textwidth]{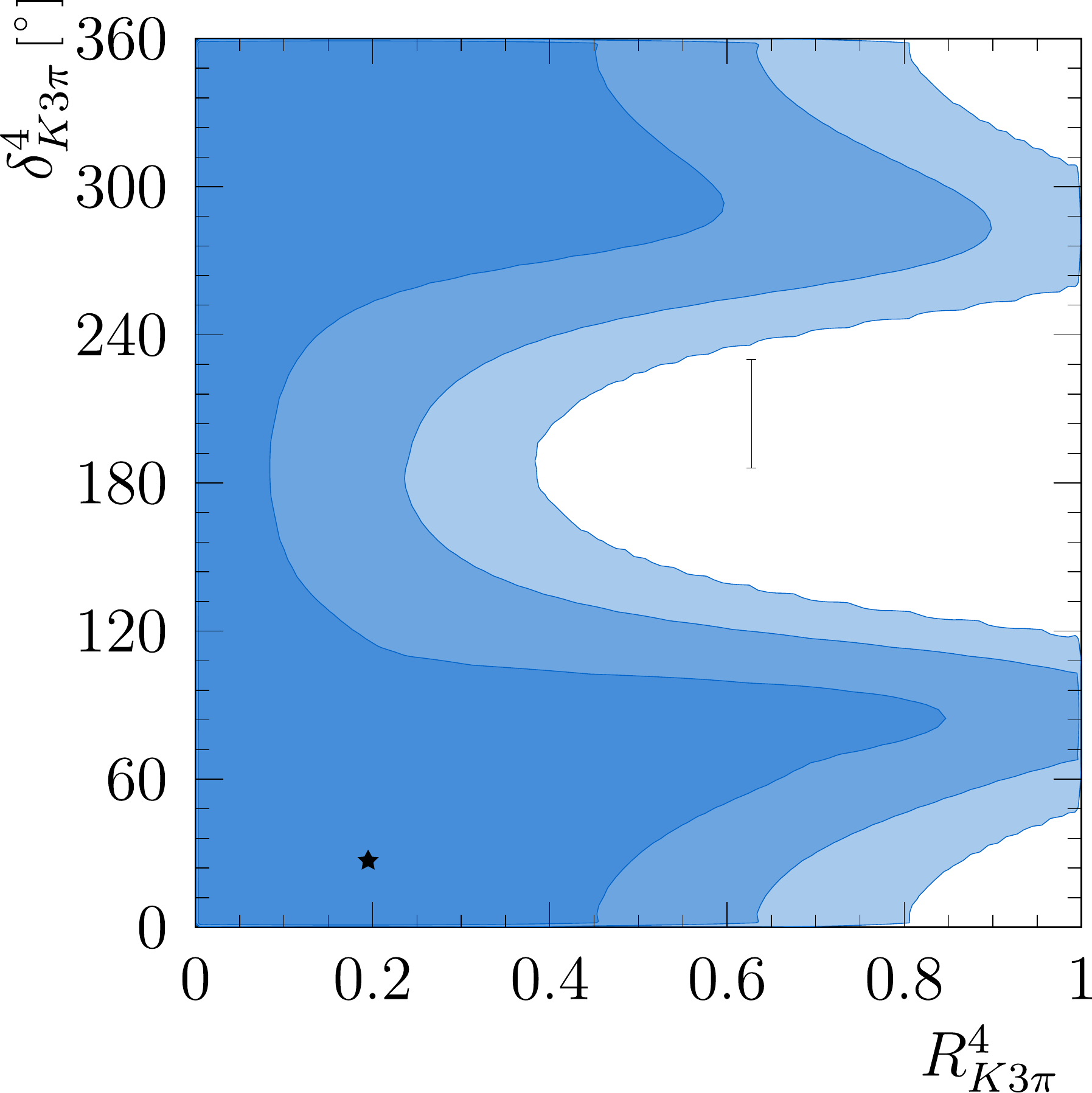}

  \caption{\label{fig:fig3}Likelihood scans for the hadronic parameters in the four bins, where the contours give the regions corresponding to $-2\ln L = 2.30, 6.18, 11.83$.  Stars mark the best fit values.
    Predictions are shown by the vertical error bars, the size of which indicates the uncertainty on the absolute phase difference measured in the global analysis~\cite{EVANS}. 
  }
\end{figure}

\begin{table}
  \centering
  \caption{\label{tb:results} Predicted and measured values of the coherence factors and strong-phase differences in  four bins of phase space. 
The predicted values are derived from the amplitude models. The absolute phase differences are obtained by adding  the global strong phase-difference measured in Ref.~\cite{EVANS} to the normalised phase difference predicted by the model.  
  }
  \vspace{1mm}  
  \begin{tabular}{l 
r
r 
r 
r
r 
}
\toprule
  & & \multicolumn{2}{c}{Predicted} & \multicolumn{2}{c}{Measured}\\
  Bin & Limits $(\tilde{\delta}_{K3\pi})$ & $R^{i}_{K3\pi}$ 
& $\delta^i_{K3\pi}$ 
    & \multicolumn{1}{c}{$R^{i}_{K3\pi}$}
    & \multicolumn{1}{c}{${\delta^{i}_{K3\pi}}$}\\
\midrule
1 & $-180^\circ < \tilde{\delta}_{K3\pi} \leq 39 ^\circ $ & 0.67  &  $56 ^\circ$ & $0.61  ^{+0.28}_{-0.54}$ & $\left(100^{+55}_{-18}\right)^\circ $ \\
  2 & $39  ^\circ < \tilde{\delta}_{K3\pi} \leq 0  ^\circ $ & 0.85  &  $108^\circ$ & $1.00  ^{+0.00}_{-0.40}$ & $\left(131^{+34}_{-12}\right)^\circ $ \\
  3 & $0   ^\circ < \tilde{\delta}_{K3\pi} \leq 43 ^\circ $ & 0.82  &  $149^\circ$ & $0.53  ^{+0.34}_{-0.21}$ & $\left(157^{+77}_{-36}\right)^\circ $ \\
  4 & $43  ^\circ < \tilde{\delta}_{K3\pi} \leq 180^\circ $ & 0.63  &  $208^\circ$ & $0.19  ^{+0.32}_{-0.18}$ & $\left( 26^{+67}_{-90}\right)^\circ $ \\
\bottomrule
\end{tabular}
 \end{table}

The fit converges with a $\chi^2$ of $166$ for $162$ degrees of freedom.
Likelihood scans are presented in  Fig.~\ref{fig:fig3} in the $R^i_{K3\pi}$~vs.~$\delta^{i}_{K3\pi}$ plane for each bin.   Also included in the plots are the predictions from the model.  The phase differences for these predictions are obtained by adding $(128^{+28}_{-17})^\circ$, the value of the measured strong-phase difference averaged over all phase space~\cite{EVANS}, to the value of the normalised phase difference calculated from the model. 
Table~\ref{tb:results} shows the numerical results, where the values for the measurements correspond to the best fit points from the scans.
The correlation matrix for these results can be found in the appendix. 
From Table~\ref{tb:results}, and the contours in Fig.~\ref{fig:fig3}, it can be seen that bins 1 to 3 show reasonable compatibility between the measurements and the predictions, but the agreement is less satisfactory in bin 4.
In order to assess the probability of such a configuration of results, under the assumption that the model correctly describes nature, an ensemble of simulated data sets is generated and fitted.
This exercise returns a $p$-value of $6\%$. 
If future studies reveal that the true value of the global phase difference is somewhat higher than the central value measured in Ref.~\cite{EVANS} then the compatibility will improve.
For example, a value of $180^\circ$, which lies within the two-sigma contour of the current measurement, would lead to a $p$-value of $28\%$. 

The BESIII collaboration have collected a $\psi(3770)$ data set approximately four times larger than that of CLEO-c.  Analysis of these data will provide a more precise test of whether the LHCb models provide a good description of the phase variation of $D \to \PIKPIPI$ decays. Again, it must be emphasised that any imperfection in the models will not bias the measurement of the CKM angle $\gamma$ with this method. 

\subsection{Impact on $\gamma$ determination}

\begin{figure}
  \centering 
  \subfloat[S][]{\label{fig4a}\includegraphics[width=0.495\textwidth]{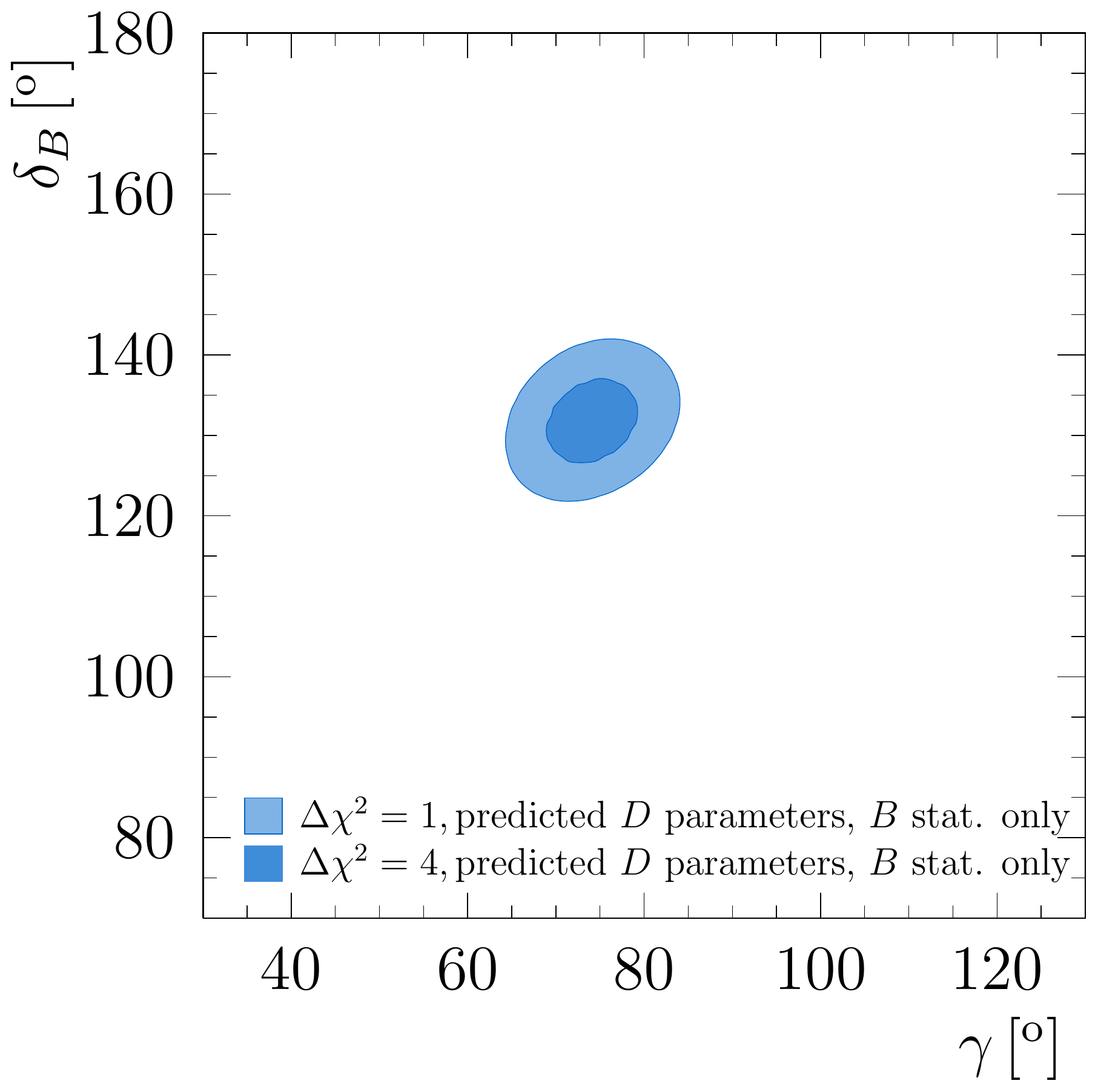}}
  \subfloat[S][]{\label{fig4b}\includegraphics[width=0.495\textwidth]{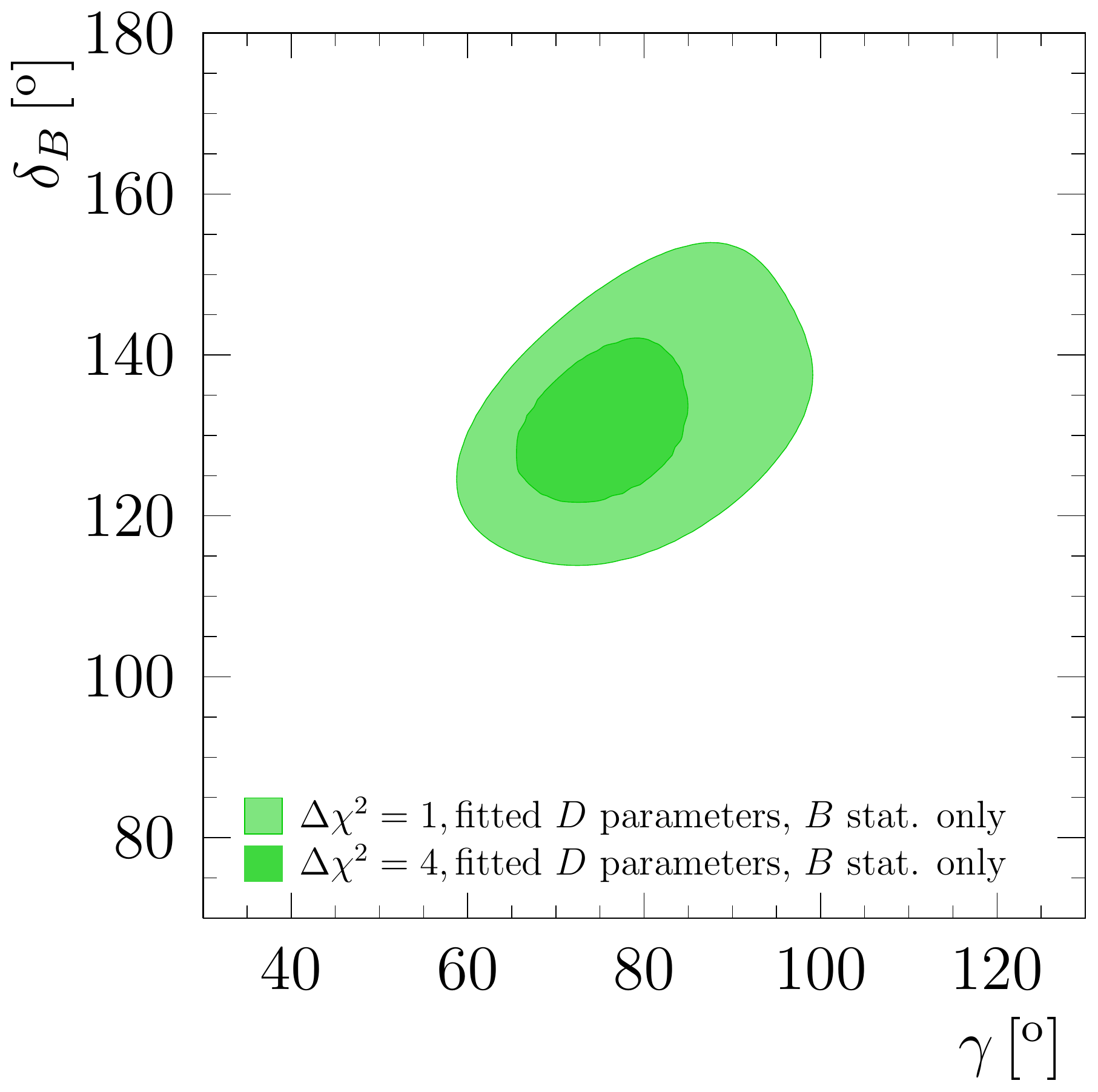}}

  \subfloat[S][]{\label{fig4c}\includegraphics[width=0.495\textwidth]{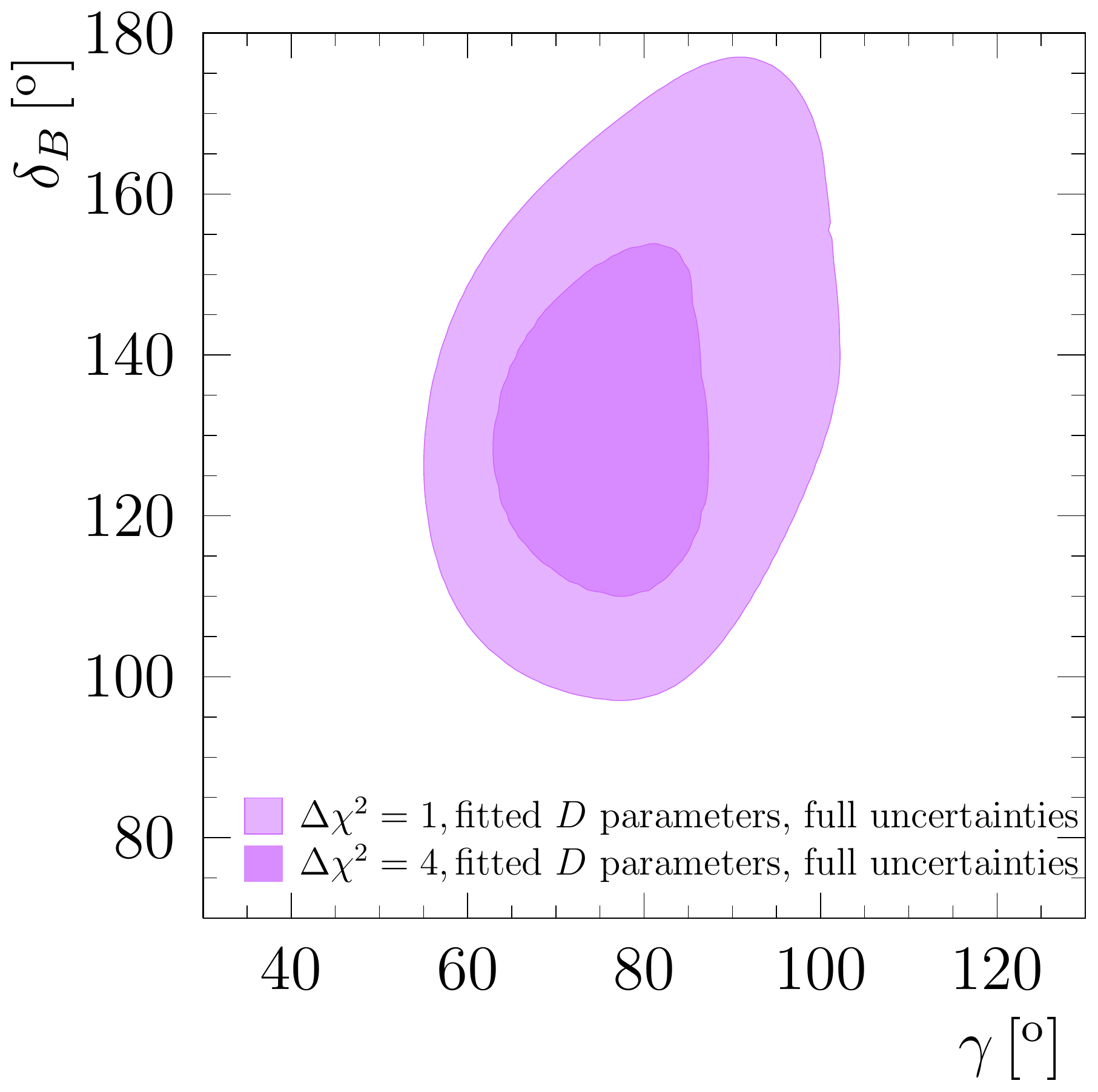}}

  \caption{\label{fig:fig4}The contours of $\Delta \chi^2$ in the $\gamma$ vs. $\delta_B$ plane, corresponding to $\Delta \chi^2 = 1,4$. Indicated is: (a) perfect knowledge of the hadronic parameters taking the predicted values, (b) perfect knowledge of the hadronic parameters, taking as central values the measurements presented in Table~\ref{tb:results}, and (c) taking the central values of the measurements in Table~\ref{tb:results} and allowing the hadronic parameters to vary within the measured uncertainties.} 
\end{figure}

It is of interest to understand how both the central values and the uncertainties of the measurement of the hadronic parameters from the CLEO-c data set affect the sensitivity to $\gamma$ and auxiliary parameters, $\delta_B$ and $r_B$.   Therefore the simulation studies described in Sec.~\ref{subsec:toys} are  extended to address these questions for the four-bin scheme.  Three scenarios are considered and shown in Fig.~4: (a) assuming perfect knowledge of the hadronic parameters and taking the central values from the predictions; (b) assuming perfect knowledge of the hadronic parameters and taking the central values from the measurements in; (c) taking both the central values and uncertainties of the hadronic parameters from the measurements. In the latter case, the full likelihood from the CLEO data is used when constraining the hadronic parameters owing to the highly non-gaussian behaviour of the uncertainties. 

The contours of $\Delta \chi^2$ from this exercise are shown in Fig.~\ref{fig:fig4} in the $\gamma$ vs. $\delta_B$ plane. The widths of the one-sigma contours projected onto the $\gamma$ axis are around 4.9$^\circ$, 9.2$^\circ$ and 10.2$^\circ$ for scenarios (a), (b) and (c), respectively.  Hence, it can be seen that the central values as obtained from the analysis of CLEO-c data would lead to a significant degradation in sensitivity, in particular through smaller values of some of the coherence factors and reduced separation between the central values of the strong-phase differences. Nonetheless, the results from this data set would still allow for a measurement that is intrinsically more precise than is provided in a global analysis of the $D \to \PIKPIPI$ phase space.  
The uncertainty on $\gamma$ associated with the finite size of the CLEO-c data set is estimated to be around 4.4$^\circ$, by  subtracting the results of scenarios (c) and (b) in quadrature.  Analysis of the existing $\psi(3770)$ sample collected by BESIII, and complementary results from updated $D\to\PIKPIPI$ mixing studies at LHCb, should allow for this uncertainty to be halved, which would make it sub-dominant for the LHCb Run 1 and 2 data sets, even for the case of the hadronic parameters taking the values predicted by the model.  More data taking by BESIII at the $\psi(3770)$ resonance or by a future super tau-charm factory~\cite{STCF} would lead to further improvement.
 
\section{Conclusions}
\label{sec:conclusions}
All previous analyses of $B^- \to DK^-$, $D \to \PIKPIPI$ decays have integrated over the phase space of the $D$ meson, an approach which has limited the sensitivity that the measurement provides to the angle $\gamma$ of the unitarity triangle. By making use of amplitude models of the $D$ decays recently reported by the LHCb collaboration~\cite{LHCBK3PI},  it is possible to define bins in phase space that have high coherence, which will in turn allow for significantly increased sensitivity to $\gamma$. Assuming that a four-bin scheme optimised on the models has similar sensitivity when applied in data, and with perfect knowledge of the hadronic parameters of the $D$ decays, it is estimated that an uncertainty of around $5^\circ$ will be attainable with the $B$-meson data set currently available at LHCb. This precision is similar to that expected from $B^- \to D K^-, D\to\KS h^+ h^- \left[h=\pi,K\right]$ decays, which currently provide the best standalone sensitivity to $\gamma$ \cite{LHCBGAMMAKSHH}.
These bin definitions will also be valuable for studies of mixing and $\CP$ violation in the $D^0\bar{D}^{0}$ system. 

Data collected by the CLEO-c experiment at the $\psi(3770)$ resonance have been analysed to obtain constraints  on the hadronic parameters of the $D$ decay, which can be compared with the predictions derived from the models.  Broad consistency is observed, albeit with some tension seen in one bin.   Analysis of the larger $\psi(3770)$ sample obtained by the BESIII collaboration, and complementary studies of charm mixing above threshold, will be necessary to make a stronger test of the model predictions and provide sufficiently precise inputs for the exploitation of the LHCb Run 1 and 2 data.  Matching the excellent statistical precision that can be expected in the coming decade from the LHCb Upgrade~I~\cite{UPGRADEI} and Belle~II~\cite{BELLEII}, and beyond with LHCb Upgrade~II~\cite{UPGRADEII}, will require BESIII to collect more $\psi(3770)$ data and also provides excellent motivation for the construction of a super tau-charm factory. 
 
\section*{Acknowledgments}

This Letter includes the results of an analysis performed using CLEO-c data. The authors (some of whom were members of CLEO) are grateful to the collaboration for the privilege of using these data. 
We are grateful for support from 
the UK Science and Technology Facilities Council, the UK India and Education Research Initiative and the European Research Council under Horizon 2020.

\appendix

  \section*{Appendix A}
  \renewcommand{\thesubsection}{A.1} 
  \setcounter{table}{1}
  \renewcommand{\thetable}{\arabic{table}}
\subsection{Measured values of the double-tag observables in the CLEO-c data}
  The $\Delta^{K3\pi}_{\CP}$, $\rho^{K3\pi}_{K\pi,LS}$ and $\rho^{K3\pi}_{K\pi\pi^0, LS}$ observables, defined in Ref.~\cite{EVANS}, 
are convenient for conveying information about the double-tag yields involving the $\CP$ and the like-signed-kaon tags. 
Any deviation of $\Delta^{K3\pi}_{\CP}$ from zero, or the $\rho^{K3\pi}_{K\pi(\pi^0), LS}$ observables from one, is indicative of a non-zero coherence factor. 
These quantities are given here for the current analysis. In the fit presented in this Letter, however, it is the yields for these double tags that are used as input, due to the limitations of using the normal approximation of uncertainties for such small numbers of candidates.
The measured values in each of the four bins defined in this paper are given in Table~\ref{tb:rhoValues}, and the correlation matrix for these measurements is shown in Table~\ref{tb:corMatrix}. 
Similarly, the 10 $\rho^{K3\pi}_{LS}$ observables in the four-by-four bins for the self tags are given in Table~\ref{tb:k3piValues}.
The correlations between the results in these bins is negligible. 
Finally, the background-subtracted yields of the self-conjugate tags, for the sixteen bins of $D\to K^0_{\rm{S}} \pi^+\pi^-$ phase space and the four bins for the $D\to K^+\pi^-\pi^-\pi^+$ decay are given in Table~\ref{tb:kspipiValues}. 
The yields are corrected for relative efficiency variations, and the correlations between the results in these bins are negligible. 
\begin{table}
  \centering
  \caption{\label{tb:rhoValues}Central values and uncertainties, both statistical and systematic, for the $\Delta^{K3\pi}_{\CP}$ and $\rho^{K3\pi}_{K\pi(\pi^0),LS}$ observables in the four bins of the $\Dz\to\KPIPIPI$ phase space.}
  \begin{tabular}{r
    >{\collectcell\mathhack}r<{\endcollectcell} @{${}\pm{}$} >{\collectcell\mathhack}r<{\endcollectcell}
    >{\collectcell\mathhack}r<{\endcollectcell} @{${}\pm{}$} >{\collectcell\mathhack}r<{\endcollectcell}
    >{\collectcell\mathhack}r<{\endcollectcell} @{${}\pm{}$} >{\collectcell\mathhack}r<{\endcollectcell}
    >{\collectcell\mathhack}r<{\endcollectcell} @{${}\pm{}$} >{\collectcell\mathhack}r<{\endcollectcell}
    }
    \toprule
    & \multicolumn{2}{c}{Bin 1} & \multicolumn{2}{c}{Bin 2} & \multicolumn{2}{c}{Bin 3} & \multicolumn{2}{c}{Bin 4} \\
    \midrule
    $\Delta^{K3\pi}_{\CP}$        & 0.06 & 0.03  & 0.07 & 0.04  & 0.07 & 0.03 & 0.02 & 0.03 \\
    $\rho^{K3\pi}_{K\pi,LS}$      & 0.89 & 0.33  & 0.34 & 0.20  & 0.47 & 0.24 & 1.46 & 0.44 \\
    $\rho^{K3\pi}_{K\pi\pi^0,LS}$ & 1.47 & 0.34  & 0.58 & 0.25  & 0.84 & 0.26 & 1.00 & 0.28 \\
    \bottomrule
  \end{tabular}
\end{table}

\begin{table}
  \caption{\label{tb:corMatrix}Correlation matrix between the $\Delta^{K3\pi}_{\CP}$ and $\rho^{K3\pi}_{K\pi(\pi^0),LS}$ observables in the four bins of the $\Dz\to\KPIPIPI$ phase space.}

  \centering \scalebox{0.7}{
    \begin{tabular}{ r | rrrr | rrrr | rrrr }
      \toprule
      & \multicolumn{4}{c}{$\Delta^{K3\pi}_{\CP}$} & \multicolumn{4}{c}{$\rho^{K3\pi}_{K\pi,LS}$} & \multicolumn{4}{c}{$\rho^{K3\pi}_{K\pi\pi^0, LS}$} \\
      \midrule
      \multirow{4}{*}{$\Delta^{K3\pi}_{\CP}$} 
      & 1.000    & 0.296    & 0.280    & 0.309    & -0.072   & -0.043   & -0.049   & -0.101   & 0.000    & 0.000    & 0.000    & 0.000   \\
      &          & 1.000    & 0.281    & 0.310    & -0.070   & -0.042   & -0.048   & -0.098   & 0.000    & 0.000    & 0.000    & 0.000   \\
      &          &          & 1.000    & 0.309    & -0.069   & -0.041   & -0.047   & -0.097   & 0.000    & 0.000    & 0.000    & 0.000   \\
      &          &          &          & 1.000    & -0.076   & -0.046   & -0.052   & -0.107   & 0.000    & 0.000    & 0.000    & 0.000   \\
      \midrule
      \multirow{4}{*}{$\rho^{K3\pi}_{K\pi,LS}$} 
      &          &          &          &          & 1.000    & 0.048    & 0.054    & 0.111    & 0.003    & 0.002    & 0.002    & 0.002   \\
      &          &          &          &          &          & 1.000    & 0.032    & 0.066    & 0.002    & 0.001    & 0.002    & 0.001   \\
      &          &          &          &          &          &          & 1.000    & 0.076    & 0.002    & 0.001    & 0.002    & 0.002   \\
      &          &          &          &          &          &          &          & 1.000    & 0.003    & 0.002    & 0.003    & 0.002   \\
      \midrule 
      \multirow{4}{*}{$\rho^{K3\pi}_{K\pi\pi^0,LS}$} 
      &          &          &          &          &          &          &          &          & 1.000    & 0.007    & 0.010    & 0.011   \\
      &          &          &          &          &          &          &          &          &          & 1.000    & 0.005    & 0.006   \\
      &          &          &          &          &          &          &          &          &          &          & 1.000    & 0.008   \\
      &          &          &          &          &          &          &          &          &          &          &          & 1.000   \\
      \bottomrule

    \end{tabular}
  }
\end{table}

\begin{table}
  \centering
  \caption{\label{tb:k3piValues}Central values and uncertainties, both statistical and systematic, for the $\rho^{K3\pi}_{LS}$ observables. 
  The decay is symmetric under exchange of the $D$ mesons, and hence the observables are `folded' for the off-diagonal terms. }
  \begin{tabular}{r | 
    >{\collectcell\mathhack}r<{\endcollectcell} @{${}\pm{}$} >{\collectcell\mathhack}r<{\endcollectcell}
    >{\collectcell\mathhack}r<{\endcollectcell} @{${}\pm{}$} >{\collectcell\mathhack}r<{\endcollectcell}
    >{\collectcell\mathhack}r<{\endcollectcell} @{${}\pm{}$} >{\collectcell\mathhack}r<{\endcollectcell}
    >{\collectcell\mathhack}r<{\endcollectcell} @{${}\pm{}$} >{\collectcell\mathhack}r<{\endcollectcell}
    }  
    \toprule
          & \multicolumn{2}{c}{Bin 1} & \multicolumn{2}{c}{Bin 2} & \multicolumn{2}{c}{Bin 3} & \multicolumn{2}{c}{Bin 4} \\
       \midrule
    Bin 1 & 0.65  & 0.65         & 0.70  & 0.50         & 0.69  & 0.48         & 1.51  & 0.67 \\
    Bin 2 & \multicolumn{2}{c}{} & 0.00  & 0.76         & 0.00  & 0.37         & 0.96  & 0.56 \\
    Bin 3 & \multicolumn{2}{c}{} & \multicolumn{2}{c}{} & 0.71  & 0.71         & 0.93  & 0.53 \\
    Bin 4 & \multicolumn{2}{c}{} & \multicolumn{2}{c}{} & \multicolumn{2}{c}{} & 1.25  & 0.88 \\  
  \bottomrule
  \end{tabular}
\end{table}

\begin{table}
  \centering
  \caption{\label{tb:kspipiValues}
  Background-subtracted yields in the 16 bins of the
  $D\to K^0_{\rm{S}} \pi^+\pi^-$ phase space and four bins of the $D\to\PIKPIPI$ phase space, 
  with the combined statistical and systematic uncertainties. 
  The yields are corrected for relative bin-to-bin efficiency variations. 
  }
  \begin{tabular}{r | 
    >{\collectcell\mathhack}r<{\endcollectcell} @{${}\pm{}$} >{\collectcell\mathhack}r<{\endcollectcell}
    >{\collectcell\mathhack}r<{\endcollectcell} @{${}\pm{}$} >{\collectcell\mathhack}r<{\endcollectcell}
    >{\collectcell\mathhack}r<{\endcollectcell} @{${}\pm{}$} >{\collectcell\mathhack}r<{\endcollectcell}
    >{\collectcell\mathhack}r<{\endcollectcell} @{${}\pm{}$} >{\collectcell\mathhack}r<{\endcollectcell}
    }  
    \toprule
          & \multicolumn{2}{c}{Bin 1} & \multicolumn{2}{c}{Bin 2} & \multicolumn{2}{c}{Bin 3} & \multicolumn{2}{c}{Bin 4} \\
       \midrule
 1 & 78.8 & 9.5 & 72.0 & 8.7 & 90.3 & 9.9 & 105.1 & 11.0\\
 2 & 56.3 & 7.8 & 40.9 & 6.7 & 47.4 & 7.3 & 48.6 & 7.3\\
 3 & 42.6 & 6.5 & 33.2 & 6.0 & 33.9 & 6.0 & 50.7 & 7.3\\
 4 & 16.8 & 4.4 & 12.0 & 3.5 & 10.5 & 3.4 & 12.4 & 3.7\\
 5 & 46.8 & 7.1 & 32.1 & 5.8 & 31.5 & 6.1 & 51.7 & 7.8\\
 6 & 31.8 & 5.8 & 28.9 & 5.5 & 19.2 & 4.9 & 28.0 & 5.4\\
 7 & 78.9 & 9.1 & 50.6 & 7.3 & 69.6 & 8.6 & 86.8 & 9.7\\
 8 & 78.6 & 9.1 & 57.0 & 8.0 & 48.3 & 7.2 & 96.5 & 10.1\\
-1 & 45.1 & 7.4 & 36.3 & 6.5 & 53.2 & 7.8 & 44.8 & 7.1\\
-2 & 16.4 & 4.4 & 13.7 & 3.7 & 12.0 & 3.7 & 11.3 & 4.0\\
-3 & 5.2 & 2.5 & 11.6 & 3.5 & 13.9 & 3.7 & 17.7 & 4.2\\
-4 & 12.1 & 3.5 & 5.5 & 2.6 & 6.4 & 2.8 & 12.1 & 3.5\\
-5 & 25.9 & 5.3 & 19.3 & 4.7 & 23.4 & 5.1 & 25.8 & 5.9\\
-6 & 7.2 & 3.3 & 7.2 & 2.6 & 7.2 & 3.3 & 11.8 & 4.1\\
-7 & 7.2 & 2.7 & 8.9 & 3.4 & 1.3 & 1.7 & 16.1 & 4.1\\
-8 & 24.4 & 5.1 & 22.3 & 5.1 & 17.2 & 4.5 & 16.2 & 4.4\\
  \bottomrule
  \end{tabular}
\end{table}
   \renewcommand{\thesubsection}{A.2} 
  \subsection{Correlation matrix for the hadronic parameters}
  The correlation matrix for the measurement of the $D \to \PIKPIPI$ coherence factors and relative strong-phase differences found in Table.~1 is presented in Table.~\ref{tb:correlationMatrix}.

\begin{table}[ht]
\centering
\caption{\label{tb:correlationMatrix}Correlation matrix for the $D \to \PIKPIPI$ coherence factors and relative strong-phase differences}
\begin{tabular}{l | 
r
r 
r 
r
r
r
r
r
}
\toprule
      &	$R^{1}_{K3\pi}$	&   $\delta^{1}_{K3\pi}$  &   $R^{2}_{K3\pi}$	& $\delta^{2}_{K3\pi}$	& $R^{3}_{K3\pi}$ & $\delta^{3}_{K3\pi}$  &	$R^{4}_{K3\pi}$	& $\delta^{4}_{K3\pi}$	\\
\midrule
$R^{1}_{K3\pi}$	&	1.000	&	-0.082	&	0.223	&	-0.058	&	0.130	&	-0.180	&	0.098	&	-0.026	\\
$\delta^{1}_{K3\pi}$	&		&	1.000	&	-0.046	&	0.016	&	0.130	&	-0.036	&	0.257	&	0.149	\\
$R^{2}_{K3\pi}$	&		&		&	1.000	&	0.072	&	0.104	&	-0.097	&	-0.128	&	-0.123	\\
$\delta^{2}_{K3\pi}$	&		&		&		&	1.000	&	0.011	&	0.095	&	0.095	&	0.029	\\
$R^{3}_{K3\pi}$	&		&		&		&		&	1.000	&	-0.814	&	0.388	&	0.530	\\
$\delta^{3}_{K3\pi}$	&		&		&		&		&		&	1.000	&	-0.398	&	-0.530	\\
$R^{4}_{K3\pi}$	&		&		&		&		&		&		&	1.000	&	0.307	\\
$\delta^{4}_{K3\pi}$	&		&		&		&		&		&		&		&	1.000	\\

\bottomrule
\end{tabular}
 \end{table}
 
\FloatBarrier

\end{document}